\documentclass[prx,twocolumn,superscriptaddress,nopacs]{revtex4}
\usepackage{graphicx}
\usepackage{amsmath}
\usepackage{amssymb}
\usepackage{bm}
\usepackage{color}
\usepackage[dvipsnames]{xcolor}
\usepackage[colorlinks=true,
                linkcolor=blue,
	        citecolor=blue,
	        urlcolor=blue]{hyperref}
\newcommand{\dtp}{Department of Theoretical Physics, University of Geneva, CH-1211 Geneva, Switzerland}

\begin{document}

\title{Spin-Orbit Coupling in Transition Metal Dichalcogenide Heterobilayer Flat Bands}
\author{Louk Rademaker}
\affiliation{\dtp}

\date{\today}

\begin{abstract}
The valence flat bands in transition metal dichalcogenide (TMD) heterobilayers are shown to exhibit strong intralayer spin-orbit coupling.
This is reflected in a simple tight-binding model with spin-dependent complex hoppings based on the continuum model.
A perpendicular electric field causes interlayer hybridization, such that the effective model is equivalent to the Kane-Mele model of topological insulators.
The proposed model can be used as a starting point to understand interactions and the experimentally observed topological phases.
\end{abstract}

\maketitle


Heterostructures with Moir\'{e} patterns, emerging from the lattice mismatch or misalignment of two-dimensional materials, have proven to be a fertile ground for novel electronic states.\cite{Andrei:2020asg,Carr:2020cm,Balents:2020.725B}
Amongst possible materials, `flat bands' are expected to occur in bilayers of transition metal dichalcogenide (TMD) monolayer semiconductors. TMD heterobilayers, when the two layers are of a different material, have been proposed as an ideal realization of the triangular lattice Hubbard model with real short-range hoppings.\cite{Wu:2018ic} 

Indeed, experimental results in WS$_2$/WSe$_2$ and MoTe$_2$/WSe$_2$ bilayers include the observation of a Mott insulating state at half-filling of the flat bands\cite{Tang:2020bb,Li:2021cd} and generalized Wigner crystals at other fillings\cite{Regan:2020fk,Xu:2020dx,Huang:2021io,Jin:2021es}. Recently, it came as quite a surprise that upon tuning a perpendicular electric field, both the quantum spin Hall effect (QSH) and quantum anomalous Hall (QAH) effect were observed in bilayer MoTe$_2$/WSe$_2$.\cite{Li:2021vy} After all, the original heterobilayer model\cite{Wu:2018ic} did not include any spin-orbit coupling, in contrast to the homobilayer model that predicted topological bandstructures.\cite{Wu:2019gm}

There is a growing literature aiming to understand and characterize TMD heterobilayers,\cite{Shih:2020ia,Weston:2020fz,Zhang:2020hk,Shabani:2021fh,Zhang:2019wk,Zhang:2021hx,Padhi:2021bs,Vitale:2021vg,Li:2021up,Li:2021cc,Naik:2020bq,Naik:2018eha,Pan:2021uh,Pan:2020cd,Pan:2020kga} where the observed topological transitions are proposed to originate in lattice relaxation\cite{Xie:2021ay} or interlayer coupling.\cite{Zhang:2021mote2,2021arXiv211101152P} However, as of yet there is no simple tight-binding model describing topological phases.

In this Letter, I reinterpret the continuum model of Ref.~\cite{Wu:2018ic}, using symmetry arguments related to the backfolding of the monolayer momenta to the Moir\'{e} mini-Brillouin zone. This directly yields strong {\em spin-orbit coupling} in the form of a complex spin-dependent hopping. The corresponding tight-binding model is equivalent to the next-nearest neighbor hopping in the Kane-Mele (KM) model. By tuning the band offset through a perpendicular electric field, the second layer comes into play and we find a full realization of the KM model with a corresponding topological transition and band inversion to a topological insulator phase. The tight-binding model derived in this Letter can serve as a starting point for studying the interplay between topology and interaction effects in TMD heterobilayers.


{\em Continuum model -- } Let me first recap the essence of the continuum model of flat bands in TMD heterobilayers.\cite{Wu:2018ic} Monolayer transition metal dichalcogenide (MX$_2$ with M=W, Mo and X = S, Se, Te) have a hexagonal lattice with $C_3$ rotational symmetry. They are semiconductors with a relatively large direct bandgap at the ${\bf K}$ and ${\bf K'}$ point. The valence band states have a large Ising spin splitting, such that states at ${\bf K}$ are spin-$\uparrow$ and the states at ${\bf K'}$ are spin-$\downarrow$. Spin ($\uparrow$ or $\downarrow$) and valley (${\bf K}$ or ${\bf K'}$) are therefore inextricably coupled.

Whenever two different TMD monolayers are combined in a heterobilayer with twist angle $\theta$, a Moir\'{e} pattern is created with lengthscale $a_M = 1/\sqrt{\frac{1}{a_1^2} +\frac{1}{a_2^2} - \frac{2 \cos \theta}{a_1 a_2}}$ where $a_{1,2}$ are the lattice constants of the monolayers. Because different TMD monolayers have different band gaps, the top of the valence band of one layer lies in the gap of the other (a type-I or type-II band alignment). Consequently, the valence band in a heterobilayer exclusively consists of states localized in a {\em single} layer.

These heterobilayer valence band states can be described, per spin/valley, by imposing a Moir\'{e} potential $V({\bf r})$ on the monolayer valence band states,
\begin{eqnarray}
	H&=& -\frac{\hbar^2 {\bf Q}^2}{2m^*} + V({\bf r}) 
	\nonumber
	\\
	V ({\bf r}) &=& \sum_{{\bf g}_j} V_j \exp \left[ i {\bf g}_j {\bf r} \right]
	\label{Eq:MoirePotential}
\end{eqnarray}
where ${\bf g}_j = \frac{4\pi}{\sqrt{3}} ( -\sin \frac{2\pi (j-1)}{6},\cos \frac{2\pi (j-1)}{6} ) $ are the six reciprocal Moir\'{e} vectors and ${\bf Q}$ is the momentum relative to the ${\bf K}/{\bf K'}$ point of the monolayer. Because $\Delta ({\bf r})$ must be real and the $C_3$ symmetry, we have $V_1 = V_3 = V_5$ and $V_2 = V_4 = V_6 = V_1^*$,\cite{Wu:2018ic,2018PhRvB..97c5306W,2017PhRvL.118n7401W} which means one can parametrize the Moir\'{e} potential using only two parameters $(V, \psi)$ such that $V_1 = V e^{ i \psi}$. For MoSe$_2$/WSe$_2$ bilayers, Ref.~\cite{Wu:2018ic} calculated $(V,\psi) = ($6.6 meV$, -94^\circ)$. This results in a topmost valence {\em flat band} with almost perfect Gaussian Wannier orbitals centered on a Moir\'{e} triangular lattice.

\begin{figure}
	\includegraphics[width=0.32\columnwidth]{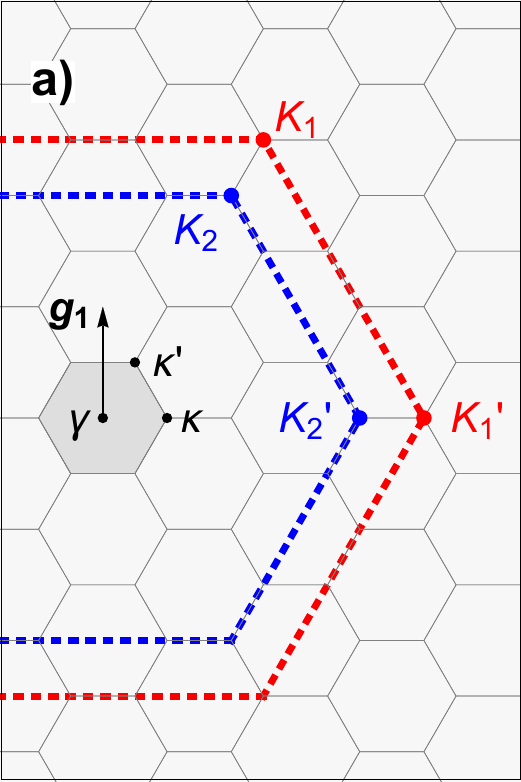}
         \includegraphics[width=0.32\columnwidth]{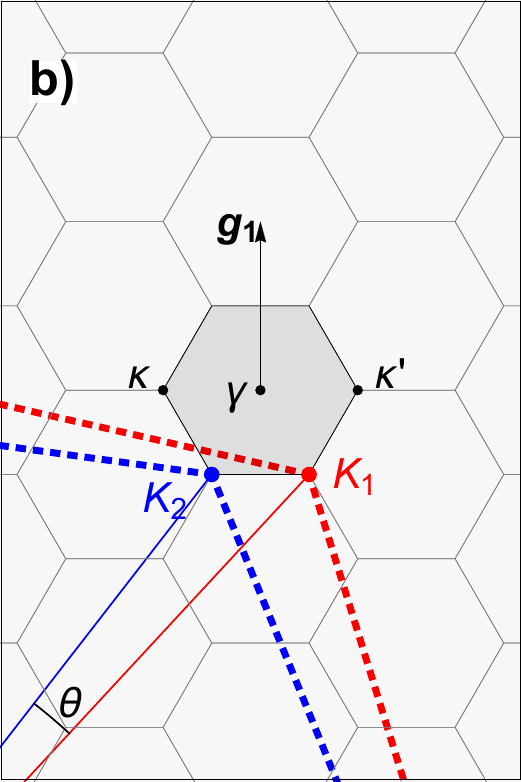}
         \includegraphics[width=0.32\columnwidth]{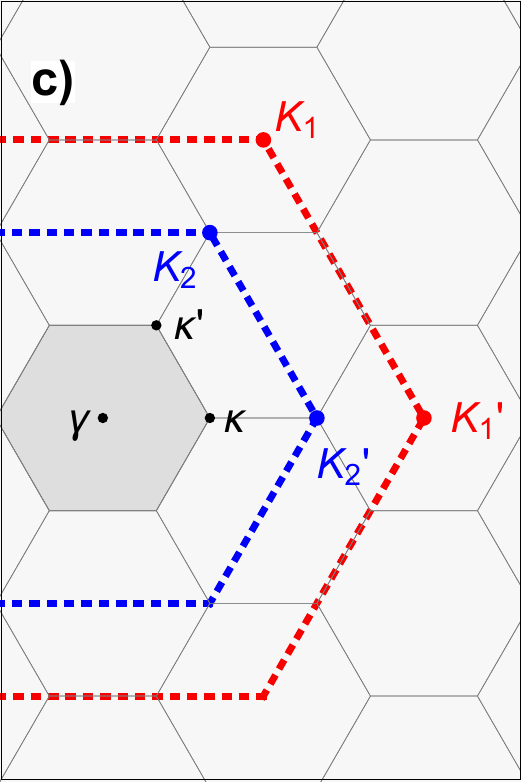}         
         	\caption{{\bf a)} When two layers with inequivalent lattice constant are combined, in general the ${\bf K}$-points of the monolayers map onto the ${\boldsymbol \kappa}$-points of the mini-Brillouin zone. This is specifically the case for aligned WS$_2$/WSe$_2$ and MoTe$_2$/WSe$_2$. {\bf b)} In case of an incommensurate Moir\'{e} pattern, for example due to a twist angle $\theta$, the natural mapping is similar. {\bf c)} Only when the Moir\'{e} length satisfies $a_M = p a_1$ with $p$ mod $3 = 0$, the ${\bf K}$-points of the layer 1 map onto the center of the mini-BZ ${\boldsymbol \gamma}$, as was done in Ref.~\cite{Wu:2018ic}.}
	\label{Fig:miniBZ}
\end{figure}

\begin{figure}[t]
	\includegraphics[width=\columnwidth]{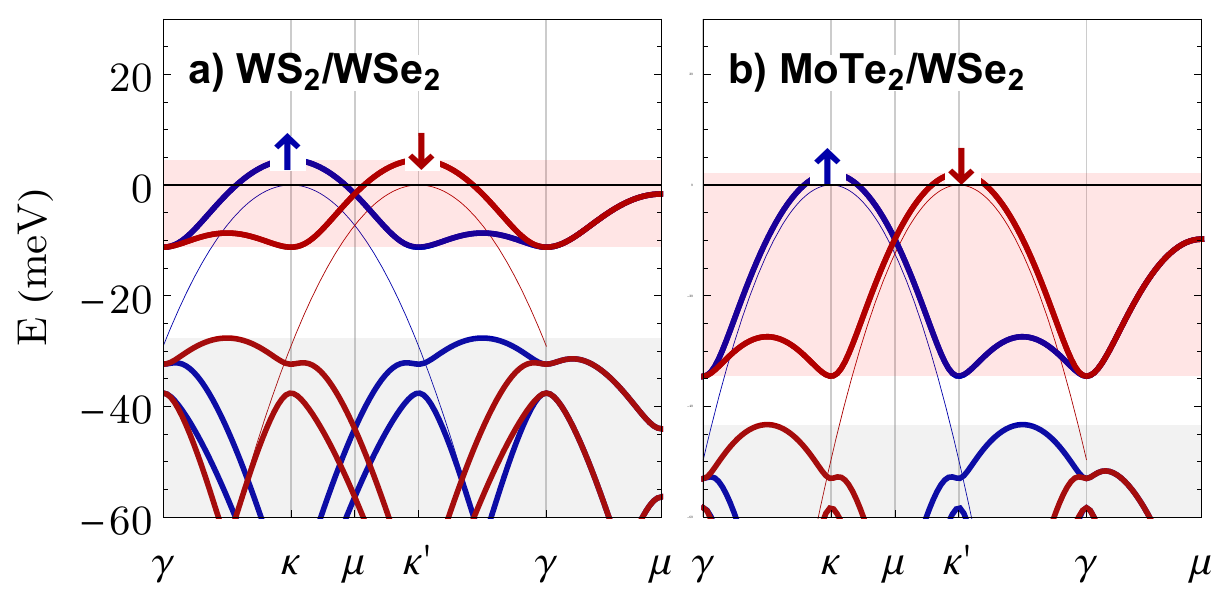}
	\caption{{\bf a)} Flat valence band structure of aligned AA-stacked WS$_2$/WSe$_2$. The colors red/blue indicate the spin $\downarrow$/$\uparrow$ of the bands, showing significant spin-orbit coupling.	
	The thin lines indicate the monolayer valence band states in the absence of the Moir\'{e} potential. Upon wannierization, the nearest neighbor hopping equals $|t_1| = 1.8$ meV. 
	{\bf b)} Flat valence band structure of aligned AB-stacked MoTe$_2$/WSe$_2$, with a similar spin-orbit splitting. The nearest neighbor hopping equals $|t_1| = 4.0$ meV.}
	\label{Fig:BandStructures2}
\end{figure}

{\em Momentum backfolding -- } The continuum model is expressed using the momenta of the monolayer, here denoted in capital letters ${\bf K}_\ell$, ${\bf K'}_\ell$ with layer index $\ell$. The Moir\'{e} potential folds the momenta back to a small mini-Brillouin zone (BZ) whose high-symmetry points are ${\boldsymbol \gamma} = (0,0)$, ${\boldsymbol \kappa} = (\frac{4 \pi}{3 a_M},0)$, ${\boldsymbol \kappa'} = ( \frac{2 \pi}{3 a_M}, \frac{2 \pi}{\sqrt{3} a_M})$ and ${\boldsymbol \mu }= \frac{1}{2} {\bf g}_1$. The question is which monolayer momenta correspond to which mini-BZ momenta.

In Ref.~\cite{Wu:2018ic}, the authors map the monolayer ${\bf K}$ to the ${\boldsymbol \gamma}$ point of the mini-BZ. Such backfolding only occurs when the Moir\'{e} length $a_M$ is a multiple of three times the monolayer lattice constant, see Fig.~\ref{Fig:miniBZ}c. In general, however, this is not true: the Moir\'{e} length in aligned commensurate cases can be expressed as $a_M = p a_1 = q a_2$ for coprime integers $p,q$. When both $p,q$ are {\em not} multiples of three, the ${\bf K}$-points of the single layers map onto the ${\boldsymbol \kappa}$-points of the mini-BZ. This is the case for aligned WS$_2$/WSe$_2$ with $a_M = 25 a_{\mathrm{WSe}_2} = 26 a_{\mathrm{WS}_2}$\cite{Li:2021cc} and aligned MoTe$_2$/WSe$_2$ with $a_M = 13 a_{\mathrm{MoTe}_2}  = 14 a_{\mathrm{WSe}_2}$.\cite{Zhang:2021mote2} Therefore the relevant momentum backfolding to describe the experiments is ${\bf K}_1, {\bf K'}_2 \rightarrow {\boldsymbol \kappa}$ and ${\bf K}_2, {\bf K'}_1 \rightarrow {\boldsymbol \kappa'}$, as is shown in Fig.~\ref{Fig:miniBZ}a. For general incommensurate and possibly twisted Moir\'{e} patterns we retain this mapping of momenta as shown in Fig.~\ref{Fig:miniBZ}b. Note that in the continuum theory of TMD homobilayers the same mapping of momenta is used.\cite{Wu:2019gm}


\begin{figure}
	\includegraphics[width=0.48\columnwidth]{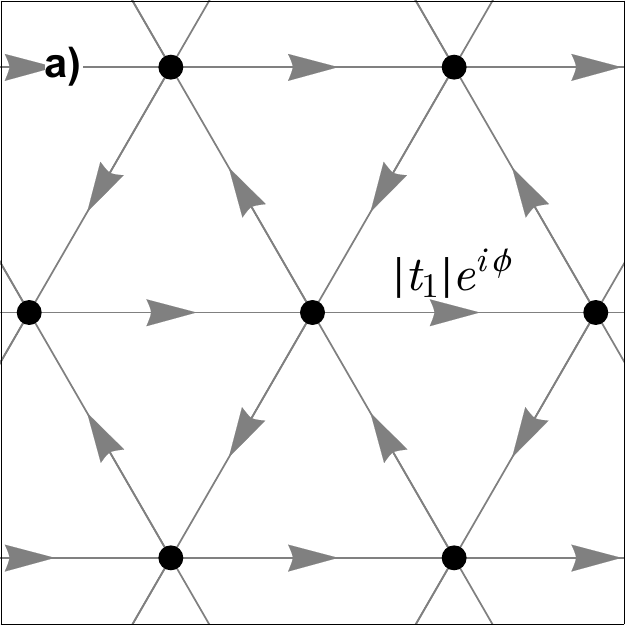}
         \includegraphics[width=0.48\columnwidth]{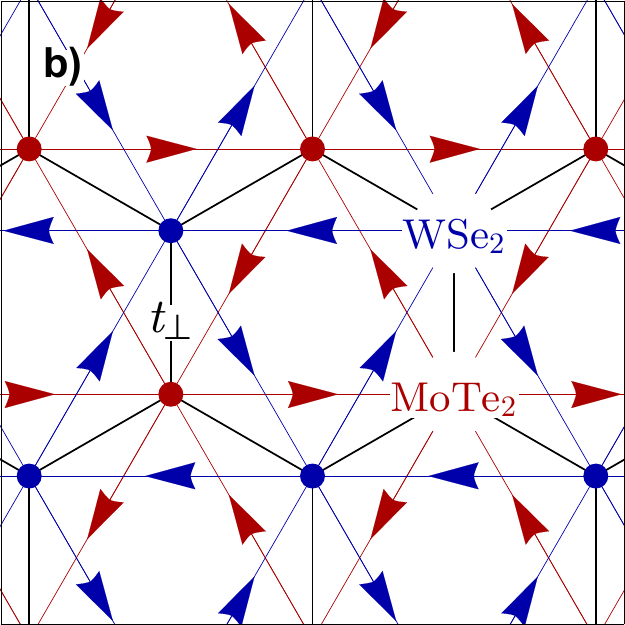}\\
	\caption{{\bf a)} The effective model of a plain heterobilayer contains orbitals at triangular lattice sites with nearest neighbor hopping $t_1 = |t_1| e^{i \phi}$ where the phase is spin-dependent, $\phi = \frac{2\pi}{3} \sigma^z$. The states live completely within one monolayer. In the case of MoTe$_2$/WSe$_2$, the states are centered at MM' sites in the MoTe$_2$ layer.
	{\bf b)} By tuning the vertical displacement field the second layer becomes relevant. The second layer, in the case of AB-stacked MoTe$_2$/WSe$_2$, is described by the same triangular lattice model with opposite spin-orbit coupling phases and a different Moir\'{e} unit cell center XX'. As a result, the combination of in-plane spin-orbit coupling Eq.~\eqref{Eq:KM1} and interlayer hopping Eq.~\eqref{Eq:Interlayer} leads to an ideal realization of the Kane-Mele model.}
	\label{Fig:LatticeModel}
\end{figure}

With the correct momentum backfolding, the valence band structure of aligned AA-stacked WS$_2$/WSe$_2$ and AB-stacked MoTe$_2$/WSe$_2$ are shown in Fig.~\ref{Fig:BandStructures2}. For WS$_2$/WSe$_2$, I use $a_M = 7.98$ nm, $m^* = 0.36 m_e$, and the Moir\'{e} potential $(V,\psi) = (7.7$ meV, $-106^\circ)$ is calculated using density functional theory with Quantum ESPRESSO\cite{2017JPCM...29T5901G,2009JPCM...21M5502G} with a Coulomb cut-off\cite{Sohier:2017fl} and the method of Ref.~\cite{Wu:2018ic}. For MoTe$_2$/WSe$_2$, I use the parameters $a_M = 4.55$ nm, $m^* = 0.65 m_e$, and $(V,\psi) = (7$ meV, $-14^\circ)$ based on Ref.~\cite{2021arXiv211101152P}. 
The resulting flat bands have a very clear spin-orbit splitting at ${\boldsymbol \kappa}$ and ${\boldsymbol \kappa'}$. The general structure of spin/valley-split bands except along the line $\gamma - \mu$ was also observed in large-scale DFT calculations of WS$_2$/WSe$_2$\cite{Li:2021cc} and MoTe$_2$/WSe$_2$\cite{Zhang:2021mote2}.

{\em Tight-binding model --} The tight-binding model for the valence flat bands can be derived using symmetry arguments. I will focus only on nearest neighbor hopping $t_1$ since it is much larger than the longer-ranged hoppings $t_2, t_3$. The $C_3$ symmetry of the heterobilayer implies that $t_1$ can be complex, $t_1 = |t_1|e^{i \phi}$, as shown in Fig.~\ref{Fig:LatticeModel}a. Within the continuum model of Eq.~\eqref{Eq:MoirePotential}, the band-structure is {\em six}-fold symmetric around the top of the flat band. Given that the top of the flat band is positioned at ${\boldsymbol \kappa}$ or ${\boldsymbol \kappa'}$, it directly follows that the complex phase equals $\phi = \pm \frac{2\pi}{3}$, respectively. The resulting tight-binding model is thus a triangular lattice with spin-orbit coupled hopping
\begin{equation}
	H =  t_1 \sum_{\langle ij \rangle \sigma} e^{i \phi \sigma^z \nu_{\langle ij \rangle}}
		c^\dagger_{i \sigma} c^{\phantom{\dagger}}_{j \sigma}
	\label{Eq:SOC}
\end{equation}
where $\nu_{\langle ij \rangle} = \pm 1$ depending on the direction of the bond, $\sigma^z = \pm 1$ indicates the $z$-component of the electron spin, and $\phi = \frac{2\pi}{3}$. What is left is to derive the magnitude of the hopping $|t_1|$, and the positioning of the triangular lattice sites, both of which depend on the specific choice of heterobilayer compounds.

Because a honeycomb lattice is equivalent to two intertwined triangular lattices, we see that Eq.~\eqref{Eq:SOC} corresponds for a single spin to the hopping model introduced by Haldane\cite{Haldane:1988gh} for the Chern insulator. Including both spins, one find that the Kane-Mele model of topological insulators,\cite{Kane:2005hl,Kane:2005gb} restricted to one honeycomb sublattice, is equal to Eq.~\eqref{Eq:SOC} with $\phi=\pi/2$.


\begin{figure}[t]
	\includegraphics[width=\columnwidth]{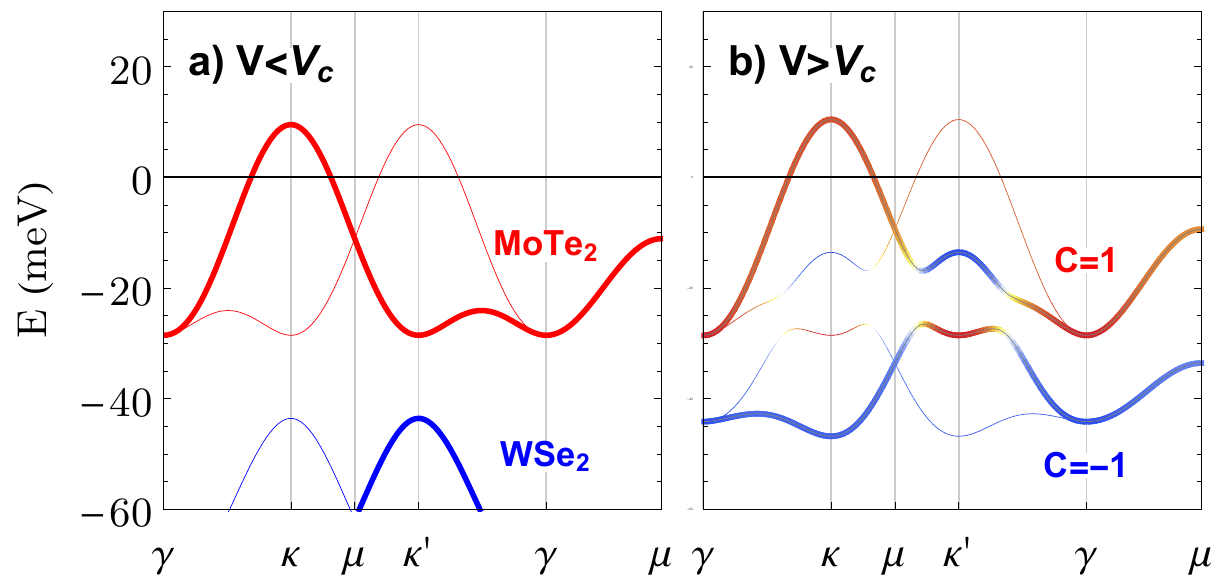}
	\caption{{\bf a.} The flat valence bands in the MoTe$_2$/WSe$_2$ heterobilayer for the ${\bf K}$-valley (the opposite valley is shown in thin lines). The top flat band contains exclusively states in the MoTe$_2$ layer, whereas the bottom band contains exclusively states in the WSe$_2$ layer.
	{\bf b.} By tuning the perpendicular electric field $V$ through a critical value, a band inversion occurs around the ${\boldsymbol \kappa}'$ point. The colorscale indicates whether states live in the MoTe$_2$ layer (red) or WSe$_2$ layer (blue). The resulting bands carry a nonzero Chern number with opposite Chern numbers for opposite spin. This transition to a topological insulator bandstructure is described by the Kane-Mele model of Eqs.~\eqref{Eq:KM1}-\eqref{Eq:Mass}.}
	\label{Fig:TopoTransition}
\end{figure}

{\em Topological transition in MoTe$_2$/WSe$_2$ -- } It has been argued that the observed topological transition in MoTe$_2$/WSe$_2$\cite{Li:2021vy} is due to a coupling to the states in the second layer,\cite{Zhang:2021mote2,2021arXiv211101152P} induced by a perpendicular electric field. There is, however, not yet a simple tight-binding model explaining this topological transition. Building on the model of Eq.~\eqref{Eq:SOC}, we can show how topologically nontrivial bands arise in a hopping model.

In the absence of an electric field, the valence flat band of MoTe$_2$/WSe$_2$ is localized in the MM' region of the MoTe$_2$ layer. In the other layer, the Moir\'{e} potential localizes the top of the valence band at the XX' region. Furthermore, because the valence band top in the ${\bf K}$-valley of WSe$_2$ maps onto ${\boldsymbol \kappa'}$, we find that the effective tight-binding model for the WSe$_2$ states has {\em opposite} phase $\phi = \mp \frac{2\pi}{3}$. The resulting two flat bands can therefore be represented by a honeycomb lattice as is shown in Fig.~\ref{Fig:LatticeModel}b. The {\em intralayer} hopping in the MoTe$_2$/WSe$_2$ bilayer is represented by the spin-orbit coupling term
\begin{equation}
	H = \sum_{\langle \langle ij \rangle \rangle \ell \sigma }
		t_\ell e^{i \ell \sigma^z \nu_{\langle \langle ij \rangle \rangle} \phi}
		c^\dagger_{i \ell \sigma} c^{\phantom{\dagger}}_{j \ell \sigma}
	\label{Eq:KM1}
\end{equation}
where $\langle \langle ij \rangle \rangle$ represents the next-nearest neighbor on the honeycomb lattice where $\nu_{\langle \langle ij \rangle \rangle} = \pm1$ depends on the direction, $\ell = \pm 1$ the layer index, and $\sigma^z = \pm 1$ the spin. This is exactly the Kane-Mele spin-orbit coupling term.\cite{Kane:2005hl,Kane:2005gb}

The interlayer hopping is now given by {\em nearest neighbor} hopping on the honeycomb lattice. The $C_3$ symmetry once again constrains the possible complex phases of the interlayer hopping, as described in Ref.~\cite{Zhang:2021mote2}. For the tight-binding model this yields
\begin{equation}
	H_{\perp} = t_\perp \sum_{\langle i j \rangle \sigma} e^{i \frac{2\pi}{3} \nu_{\langle i j \rangle} } c^\dagger_{i \ell \sigma} c^{\phantom{\dagger}}_{j \overline{\ell} \sigma}
	\label{Eq:Interlayer}
\end{equation}
where $\langle i j \rangle$ now couples the nearest neighbors on the honeycomb lattice. The parameter $\nu_{\langle i j \rangle}  = 0,1$ or $2$ depends on the direction and increases counterclockwise when going around the MoTe$_2$ lattice sites. Eqs.~\eqref{Eq:KM1}-\eqref{Eq:Interlayer} form an effective tight-binding model constrained by symmetry and is visualized in Fig.~\ref{Fig:LatticeModel}b.

In the absence of an electric field the flat bands from the WSe$_2$ layer are at a much lower energy than the MoTe$_2$ states. A perpendicular electric field $V$ can shift the WSe$_2$ states upward,
\begin{equation}
	H_V = (V + \Delta) \sum_{i} n_{i, \ell=2}
	\label{Eq:Mass}
\end{equation}
where $\Delta<0$ is the band offset. By increasing $V$ the bands from the two layers will overlap, at which point the interlayer coupling Eq.~\eqref{Eq:Interlayer} becomes relevant. 
With $t_2 = 3.4$ meV extracted from the effective mass of the WSe$_2$ valence band and $t_\perp = 4$ meV, we calculate the bandstructure as a function of perpendicular field in Fig.~\ref{Fig:TopoTransition}. Above the critical field value $V_c$, a band inversion happens at the ${\boldsymbol \kappa'}$ point for the states in the ${\bf K}$-valley. For $V>V_c$, the top valence flat bands obtain a nonzero Chern number which is opposite for the two spin species.\cite{Fukui:2005cn} As such, the system has become a topological insulator fully described by the honeycomb lattice model of Kane-Mele.

The effective Kane-Mele model applies to all TMD heterobilayers provided the applied perpendicular electric field is sufficient to overcome the valence band offset.


{\em Interactions -- } The presence of spin-orbit coupling has some implications for the possible interacting states. In particular, at half-filling of the topmost flat band the complex hopping phases lead to an effective spin model with Dzyaloshinskii-Moriya (DM) interactions.\cite{Pan:2020kga} 
Its strength, characterized by the hopping phase $\phi = \frac{2\pi}{3}$, stabilizes an in-plane 120$^\circ$ antiferromagnetic order. 
At the same time, Chern bands at half-filling are known to be susceptible to full spin polarization leading to a Quantum Anomalous Hall (QAH) effect.\cite{Bultinck:2019wp,Rademaker:2020fr} The Mott-to-QAH transition observed in Ref.~\cite{Li:2021vy} can possibly be understood as a metamagnetic transition from in-plane N\'{e}el to Ising ferromagnetic order.

Note that within the current model, there is no band gap between the topologically nontrivial bands. However, interaction-driven renormalization of the bands can open up the gap at full filling of the top valence flat bands.\cite{2021arXiv211101152P}


{\em Conclusion --} I showed that by correctly mapping the monolayer momenta onto the mini-BZ, the effective model of TMD heterobilayers obtains a strong spin-orbit coupling as given by Eq.~\eqref{Eq:SOC}.
The interlayer coupling can be described by a Kane-Mele model following Eqs.~\eqref{Eq:KM1}-\eqref{Eq:Mass}, yielding a topological transition as a function of perpendicular electric field.

Having an effective tight-binding model is an important step towards a full understanding of possible strongly correlated phases. This Letter shows that spin-orbit coupling cannot be ignored in further studies of TMD heterobilayers.

\

I thank Johannes Motruk, Kin Fai Mak, Fengcheng Wu, Allan MacDonald, and Vladimir Dobrosavljevi\'{c} for useful discussion. I acknowledge support by Swiss National Science Foundation via an Ambizione grant PZ00P2\_174208.



%

\end{document}